\documentclass[rnote]{aa} 
\usepackage{graphicx}

\newcommand{\be}{\begin{equation}}
\newcommand{\ee}{\end{equation}}
\newcommand{\der}[2]{\frac{d{#1}}{d{#2}}}

\newcommand{\pd}[2]{\frac{{\partial}{#1}}{{\partial}{#2}}}

\newcommand{\dvrg}[1]{\mathrm{div}{#1}}

\def\gsim{\;\raise0.3ex\hbox{$>$\kern-0.95em\raise-1.1ex\hbox{$\sim$}}\;}
\def\lsim{\;\raise0.3ex\hbox{$<$\kern-0.95em\raise-1.1ex\hbox{$\sim$}}\;}

\begin{document}
   \title{Wave instabilities in an anisotropic magnetized space plasma }

    \author{N.S. Dzhalilov\inst{1,2,3},  V.D. Kuznetsov\inst{2}
         \and
          J. Staude\inst{1}
}
   \offprints{J. Staude}

\institute{Astrophysikalisches Institut Potsdam (AIP), An der
Sternwarte 16, D-14482 Potsdam, Germany,
     \and
Institute of Terrestrial Magnetism, Ionosphere and Radio Wave
Propagation of the Russian  Academy of Sciences (IZMIRAN), Troitsk
City, Moscow Region, 142190 Russia,
 \and
     Shamakhy Astrophysical Observatory of the Azerbaijan Academy of
     Sciences (ShAO), Baku Az-1000, Azerbaijan \\
     \email{JStaude@aip.de; Namig@izmiran.ru; KVD@izmiran.ru}
}

   \date{Received ... ; accepted 1 July 2008}


\abstract
{}
{We study wave instability in an collisionless, rarefied hot plasma
(e.g. solar wind or corona). We consider the anisotropy produced by
the magnetic field, when the thermal gas pressures across and along
the field become unequal.}
{We apply the 16--moment transport equations (obtained from the
Boltzmann-Vlasov kinetic equation) including the anisotropic thermal
fluxes. The general dispersion relation for the incompressible wave
modes is derived. }
{It is shown that a new, more complex wave spectrum with stable and
unstable behavior is possible, in contrast to the classic fire-hose
modes obtained in terms of the 13--moment integrated equations. }
{}

\titlerunning{Wave instabilities in an anisotropic plasma}
\authorrunning{Dzhalilov, Kuznetsov \&  Staude}

\maketitle
\section{Introduction}
An almost collisionless, rarefied, hot, magnetized space plasma such
as that of the solar corona is anisotropic and inhomogeneous, in
particular in the cross-field direction (see \cite{Asc05}). There
have been observations of a thermal anisotropy of
$T_\bot/{\mathrm{T_\|}}\!\sim\!2\!-\!3$ in the solar wind by both
\cite{Feldman74} and \cite{Marsch82}; of large heavy-ion thermal
anisotropies ($T_\bot/{\mathrm{T_\|}}\!>\!100 $) by \cite{Kohl98}
and \cite{Cran99}; of protons by \cite{Cran99}; and of the coronal
hole temperature by \cite{Dode98} and \cite{Anto00}. An opposite ion
temperature relation $T_\|>T_\bot$ is also found in solar wind
observations (see \cite{Marsch06}). Due to the anisotropy in the
kinetic temperatures of protons and heavy ions, the corresponding
partial pressures become anisotropic, in addition to the total
thermal pressure such that $p_\bot\ne p_\|$. It is now generally
accepted that the observed large ion temperature anisotropies are
related to the physical mechanism by which the solar corona and
solar wind are heated (see \cite{Hollweg02} and \cite{Marsch06}).

In these circumstances, it is difficult to develop a traditional
hydrodynamical description of the plasma. We therefore attempt to
extend the MHD approximation by considering the anisotropy of the
magnetized plasma. We consider the large-scale wave peculiarities
that can appear in a collisionless plasma. The large-scale plasma
motions are usually described by a fluid approximation, and the
integrated moment equations derived from the Boltzmann-Vlasov
kinetic equation are used. If collisions between particles are rare
and a strong magnetic field approximation is valid, the usual MHD
equations have to be replaced by other equations for which the fluid
approximation is valid too. It was shown, e.g. by \cite{Grad},
\cite{Chew}, and \cite{Rudakov}, that for a collisionless plasma --
mainly across a magnetic field -- the fluid approach can be used.
However, these so-called 13--moment equations cannot be used to
describe a plasma of arbitrary anisotropic pressure. We therefore
use the 16--moment equations.
\section{Basic equations and wave equations}
The 16--moment set of equations was used by many authors in
different theoretical approaches, especially for modeling the solar
wind (see \cite{Demars79}, \cite{Olsen99}, \cite{Li99}, and
\cite{Lie01}). The 16--moment set of transport equations for the
collisionless plasma in the presence of gravity $g$ but without
magnetic diffusivity is given as follows (see e.g. \cite{Oraev85}):
\begin{eqnarray}
&&\der{\rho}{t}+\rho\,\dvrg{\vec v}=0,  \label{rho}\\
&&\rho\der{\vec
v}{t}+\nabla(p_\perp+\frac{B^2}{8\pi})-\frac{1}{4\pi}
(\vec{B}\cdot\nabla)\vec{B}
= \rho{\vec g}+ \nonumber\\
&&+(p_\perp-p_\parallel)[{\vec h}\dvrg{\vec h}+({\vec
h}\cdot\nabla){\vec h}]+
{\vec h}({\vec h}\cdot\nabla)(p_\perp-p_\parallel), \label{vel}\\
&&\der{}{t}\frac{p_\parallel B^2}{\rho^3}=
-\frac{B^2}{\rho^3}\left[B({\vec h}\cdot\nabla)
\left(\frac{S_\parallel}{B}\right)+\frac{2S_\perp}{B} ({\vec h}
\cdot\nabla)B\right],\label{ppar}\\
&&\der{}{t}\frac{p_\perp}{\rho B}= -\frac{B}{\rho}({\vec
h}\cdot\nabla)\left(\frac{S_\perp}{B^2}\right),
\label{pper}\\
&&\der{}{t}\frac{S_\parallel B^3}{\rho^4}= -\frac{3p_\parallel B^3}
{\rho^4}({\vec h}\cdot\nabla)
\left(\frac{p_\parallel}{\rho}\right),\label{spar}\\
&&\der{}{t}\frac{S_\perp}{\rho^2}\!=\!-\frac{p_\parallel}{\rho^2}\left[({\vec
h}\!\cdot\!\nabla)\!\left(\frac{p_\perp}{\rho}\right)+
\frac{p_\perp}{\rho}\,\frac{p_\perp-p_\parallel}{p_\parallel
B}({\vec h}\!\cdot\!\nabla)B
  \right],\label{sper}\\
&&\der{\vec B}{t}+\vec{B}\dvrg{\vec
v}-(\vec{B}\cdot\nabla)\vec{v}=0,  \ \ \dvrg{\vec{B}}=0,\label{maks}
\end{eqnarray}
where $\nabla=\nabla_\parallel+\nabla_\perp, \nabla_\parallel={\vec
h}({\vec h}\cdot\nabla),$
 \be
 \der{}{t}=\pd{}{t}+(\vec{v}\cdot\nabla), \,\vec{v}=\vec{v_\parallel}+
 \vec{v_\perp}, \,{\vec h}=\frac{\vec B}{B},
\ee
and $S_\|$ and $S_\bot$ are the heat fluxes along the magnetic field
of parallel and perpendicular thermal motions. If the thermal fluxes
are neglected, $S_\bot=0$ and $S_\| = 0$, we obtain the equations
describing the laws of the change in longitudinal and transverse
thermal energy along the trajectories of the plasma (the left-hand
parts of Eqs. (\ref {ppar}) and (\ref {pper})). These so-called
``double-adiabatic'' invariants and Eqs. (\ref {rho}), (\ref {vel}),
and (\ref {maks}) also form a closed system of equations, the CGL
(Chew-Goldberger-Low) equations (see \cite{Chew}). By using the
CGL-equations, we would however obtain incomplete equations instead
of Eqs. (\ref {spar}, \ref {sper}). This is because, by deriving the
CGL equations, authors so far omitted without proof the third
moments of the distribution function and therefore the thermal
fluxes (see \cite{Chew} and \cite{Baran}). The equations derived for
the 16--moment set, in our case Eqs. (\ref {rho}--\ref {maks}),
include the thermal fluxes; they are more complete, and the CGL
equations cannot be derived from these equations as a special case.
 For simplicity, we assume that the
basic initial equilibrium state of the plasma is homogeneous ($g=0$,
and the quantities $v_0, \rho_0, p_{\bot0}, p_{\|0}, B_0, S_{\bot
0}, \mathrm{and} \ S_{\| 0}$ are constant). Equations (\ref {rho}) to
(\ref {maks}) automatically satisfy this equilibrium state with
non-zero thermal fluxes. We consider small linear perturbations of
all physical variables, for example pressure in the form $p=p_0+
p'(r, t) $, where $p'(r, t) \sim \exp i (\vec {k}\cdot\vec {r} -
\omega t) $, $ \omega $ is the wave frequency, and $k $ is the wave
number. For the perturbations, we obtain linear wave equations. Even
if we insert zero initial heat fluxes $S_{\|0}=S_{\perp0}=0$, the
perturbations of these functions will never become zero:
$S'_{\|}\ne0$, $S'_{\perp}\ne0$. Using the 16-moment equations, we
should derive more reliable results about the wave properties in an
anisotropic plasma than with the CGL equations based on the
13-moment equations.

In the presence of an external magnetic field, the initial
collisionless heat fluxes should be defined by solutions of the
kinetic equations.  We should, however, use some appropriate
estimate as a parameter. The heat flux functions should be estimated
by taking the thermal energy density of the electrons multiplied by
the particle stream speed along the magnetic field $u_0$:
$S_{\|0}\approx \frac{3}{2}n_{\mathrm{e}}k_{\mathrm{B}}T_\|
u_0\,\delta = \frac{3}{4}\delta u_0 p_\|$. Hollweg (1974, 1976)
provided some estimates of the correction parameter $\delta$
($\alpha$ in these papers) by assuming realistic shapes of electron
distribution functions and comparing the results with space
observations. We note that $\delta$ depends on the magnetic field.
In the range of $B=0.1 - 100$ G, the estimates give $\delta\approx 4
- 0.1$. In the same way, $S_{\bot0}\approx \frac{3}{4}\delta u_0
p_\bot$, and we define the parameter
$\gamma=(3/4)\delta{v_0}/{c_\|}$. We note that \cite{Marsch87}
quoted values of $\gamma$ measured in the solar wind.

We introduce dimensionless parameters and note that the indices "0"
of physical parameters are omitted for simplicity: \be
\alpha=\frac{p_\bot}{p_\|},\, \, c_\|^2=\frac{p_\|}{\rho},\,
\beta=\frac{B^2}{4\pi p_\|}=\frac{v_{\mathrm{A}}^2}{c_\|^2},\, \ee
\be \bar{S_\|}=\frac{S_\|}{p_\|c_\|},\,
\bar{S_\bot}=\frac{S_\bot}{p_\bot c_\|}, \, \, l=\cos^2\phi,\, \ee
where $\phi$ is the angle between wave vector and magnetic field,
and the indices $ \parallel $ and $ \perp $ correspond to the values
of the parameters along and across the magnetic field, respectively.
We note that our $\beta$ is defined to be inversely proportional to
the more commonly used plasma beta,
$\beta=2/\beta_{\mathrm{plasma}}$. With the parameter $\gamma$
defines above, we have $\bar{S_\|}=\bar{S_\bot}=\gamma$.

In analogy with the usual MHD equations used e.g. by \cite{somov},
there are two independent wave branches in the plasma: waves that do
not compress the plasma ($\dvrg\,{\vec v} =0 $) and waves that
compress the plasma ($ \dvrg\,{\vec v} \ne 0 $). We restrict
ourselves to the incompressible wave modes.

After inserting into the wave equations the condition of
incompressibility $ (\vec k\cdot\vec v) =0 $ and $\rho'=0$, we
obtain as usual the parametric dispersion equation. This is a
polynomial equation of 6\,th order in the frequency of oscillations.
For the parameter $Z=\omega/({c_\| k_\|}) $, this equation can be
written in the form
\begin{equation}\label{Dis0}
 c_6\,Z^6+c_5\,Z^5+c_4\,Z^4+c_3\,Z^3+c_2\,Z^2+c_1\,Z+c_0=0,
\end{equation}
\[c_0=3\, \left(1-\alpha \right)  \left[ {\alpha}^{2}(l-1)+
\alpha(1+l)+2\,(\beta-l) \right],
 \]
\[c_1= 2\,\gamma \left[ \beta(\alpha-2) +3\,\alpha(1-2\,l)
+{\alpha}^{2}(4l-3)+2\,l \right],
\]
\[c_2=4\,l+ 2\,\beta(4\,\alpha-5)+2\,{\alpha}^{2}(1+3\,l)+{\alpha}^{3}(l-1)- \]
\[ -\alpha(1+11\,l), \, \,
c_3= -2\,\gamma \left[ \beta(\alpha-2)+{\alpha}^{2}(2\,l-1)-2\,\alpha \right],\]
\[c_4= 6\,l+\alpha(l-3)-2\,l{\alpha}^{2}+2\,\beta(2-\alpha), \]
\[c_5= 2\,\gamma \left( 2\,l\alpha-2\,l-\alpha \right),\, \, \,
c_6=\alpha+l\alpha-4\,l. \] All coefficients are real and,
consequently, all solutions are real or conjugate complex. In the
usual isotropic MHD case, only Alfv\'en waves with
$\omega^2=k_\|^2v_A^2$ are present, the phase velocities of which
are equal to each other in both directions with respect to the
magnetic field. Instead of assuming that $Z^2=\beta$ in the
isotropic MHD, we determined now the 6\,th order Eq. (\ref{Dis0}) in
the anisotropic case. With the heat fluxes for which $\gamma\ne 0$,
odd nonzero coefficients $c_1,c_3,$ and $c_5$ generate wave
propagation velocities that depend on the direction of the magnetic
field.
\section{Limiting and special cases}
In this section we investigate the most important cases of Eq.
(\ref{Dis0}) which can be solved analytically.
\subsection{Parallel propagation}
In the case $l=1$ or $k=k_\|$, the six roots of Eq. (\ref{Dis0}) are
simple. The first pair of roots corresponds to a pair of stable
modes
\begin{equation}\label{s1}
\omega=\pm k_\|c_\| .
\end{equation}
The properties of these modes do not depend on the magnetic field.
They are isotropic in terms of the direction of magnetic field. This
implies that stable waves propagate along and against the magnetic
field with the same phase velocity. In these modes, $\rho'=0$,
$v'=0$, $B'=0$, $p_\|'=0$, and $S_\|'=0$. However, $p_\bot'\sim
S_\bot'\ne0$. The restoring force for these modes is therefore
$\nabla p_\bot$. Motions of particles across the magnetic field
cause a perturbation of the plasma pressure $p_\bot$, but this is
compensated by the generation of a heat flux. At the same time no
hydrodynamic motion is produced by these modes, $v=0$ as
$p_\|'=0$. We therefore have unusual thermal waves: they are stable
waves ($\mathrm{Im}(\omega)=0$) that propagate
($\mathrm{Re}(\omega)\ne 0$) with the parallel sound speed $c_\|$.
Analogous to sound waves, they are not dispersive modes. The modes
of this branch are named ``isotropic thermal'' waves.

The second pair of roots corresponds to
\begin{equation}\label{th1}
\omega/k_\|c_\|=\pm \sqrt{\alpha+\beta-1},
\end{equation}
which are conventional isotropic fire-hose waves. The waves become
unstable if $\alpha+\beta<1$ or $p_\|>p_\bot + 2p_{\mathrm{mag}}$.
The fire-hose instability therefore disappears for $\beta\ge 1$ or
$\alpha\ge1$. The dispersion relation (\ref{th1}) passes into that
for the usual isotropic Alfv\'{e}n waves $\omega^2=k_\|^2
v_{\mathrm{A}}^2$ if $\alpha=1$.

The third pair of solutions corresponds to
\begin{equation}\label{par2}
    2\frac{\omega}{k_\|c_\|}=-\gamma\pm\sqrt{\gamma^2+12\frac{1-\alpha}{2-\alpha}}.
\end{equation}
The instability condition $(\alpha-1)/(2-\alpha)>\gamma^2/12$ is
obeyed if $1<\alpha<2$. For $\gamma\ne0$, the unstable modes begin
to propagate, $\mathrm{Re}(\omega)\ne0$. For $\gamma=0$, stable
waves outside the region $1<\alpha<2$ travel along and opposite to
the direction of the magnetic field with the same phase velocity.
With $\gamma\ne0$, the stable waves for which
$(\alpha-1)/(2-\alpha)<\gamma^2/12$ become anisotropic. Retrograde
waves travel more rapidly than prograde waves. These branches are
named thermally ``anisotropic'' waves. With increasing $\gamma$, the
prograde waves propagate more slowly, but retrograde modes travel
faster. There is a crossing of these branches on the axes of
$\alpha$ for $\gamma=0$ at $V_{\mathrm{ph}}=0$, and for $\gamma>0$
at $V_{\mathrm{ph}}<0$. Between the crossing points of the branches
and $\alpha=2$, an instability arises. In contrast to the fire-hose
and the thermal modes, these modes in the instability region are
traveling, $\mathrm{Re}(\omega)\ne0$. They do not depend on $\beta$.
For these waves
\[\frac{p_\|'}{p_\|}=2(1-\alpha)\frac{B'}{B},\ \ \
 \frac{p_\bot'}{p_\bot}=\frac{1-\alpha^2}{1-2\alpha}\frac{B'}{B},\]
 \[S_\|'\sim p_\|',\ \ \ S_\bot'\sim 3\alpha\frac{1-\alpha}{1-2\alpha}\frac{B'}{B} . \]
These modes are generated by pressure anisotropy: if $\alpha=1$,
they disappear.
\subsection{Oblique propagation}
\subsubsection{Strong magnetic field}
The limit $\beta\gg1$  produces
solutions of Eq.(\ref{Dis0}) that are similar to those in the
parallel propagation case. In this case, the fire-hose waves become
stable with high phase velocities, $Z^{2}\sim O(\beta)$. Other
isotropic thermal and thermally anisotropic waves remain unchanged.\\
\subsubsection{Weak magnetic field}
 In the limit $\beta\ll1$, the first
pair of solutions of Eq.(\ref{Dis0}) simplifies to $Z^2\approx
\alpha-1$, which is the same solution as that for the parallel
fire-hose waves. The other thermal modes are described by a
polynomial equation of 4\,th order. However, for the limit
$\gamma\to0$, we obtain
\begin{equation}\label{obl4}
Z^2\approx (-b_*\pm\sqrt{D})/(2a_*),\,\,D=b_*^2-4a_*c_* ,
\end{equation}
 where $c_*=3(1-\alpha)[\alpha(1-l)-2l] ,\ \  a_*=\alpha+l(\alpha-4) ,$
and $b_*=\alpha^2(1-l)-4\alpha(1+l)+10l .$
 These solutions can easily be investigated. In the region
$0<\alpha<1$ for a closed area ($\alpha, l$), we have $D<0$ and,
therefore, $\mathrm{Re}(Z)\ne 0$ and $\mathrm{Im}(Z)\ne0$. This
implies that both thermal waves become unstable, and
 $D>0$ if $\alpha\ge 1$. In this case for unstable modes,
  $\mathrm{Re}(Z)=0$ and instability
disappears if $\alpha\ge 2$. The thermal wave branches have crossing
points, where their phase velocities coincide. For example, this
occurs if $l=1$ and $\alpha=0.5$.
\subsubsection{Isotropic propagation}
We consider the important special case of arbitrary $\beta$ and
$\gamma=0$. Even though the absence of fluxes, $S_\perp=0$ and
\,$S_\|=0$, is far from reality, this simplified case was
investigated using the 13--moment equations (e.g. \cite{Kato},
\cite{Baran}, and \cite{Kuzn}).  Substituting $\gamma=0$ into Eq.
(\ref{Dis0}), we obtain a cubic equation for $\zeta=Z^2$: \be
\label{cub} c_6\zeta^3 + c_4\zeta^2 + c_2\zeta + c_0 = 0 . \ee We
have symmetrically only three pairs of solutions,
$Z=\pm\sqrt{\zeta}$. The analytical solutions $\zeta_1,\zeta_2$, and
$\zeta_3$ allow us to investigate in significant detail the
dependence of the solutions on the parameters $\alpha,\beta$, and
$\phi$. For a small deviation of the propagation angle ($\phi\ne0$),
the situation differs significantly from that of the parallel
propagation. All three modes interact, and this interaction occurs
in two ranges of $\alpha$: $\alpha>\alpha_c$ and $\alpha<\alpha_c$.
The critical value of $\alpha=\alpha_c=4l/(1+l)$ is the  singular
point for $c_6=0$. With increasing propagation angle, the
interaction domains expand. In these domains, the growing rates also
become larger and we obtain a mixture of modes --- a turbulent wave
motion.
\subsubsection{Anisotropic propagation}
\begin{figure*}
\centering
\hfill\includegraphics[width=0.43\textwidth]{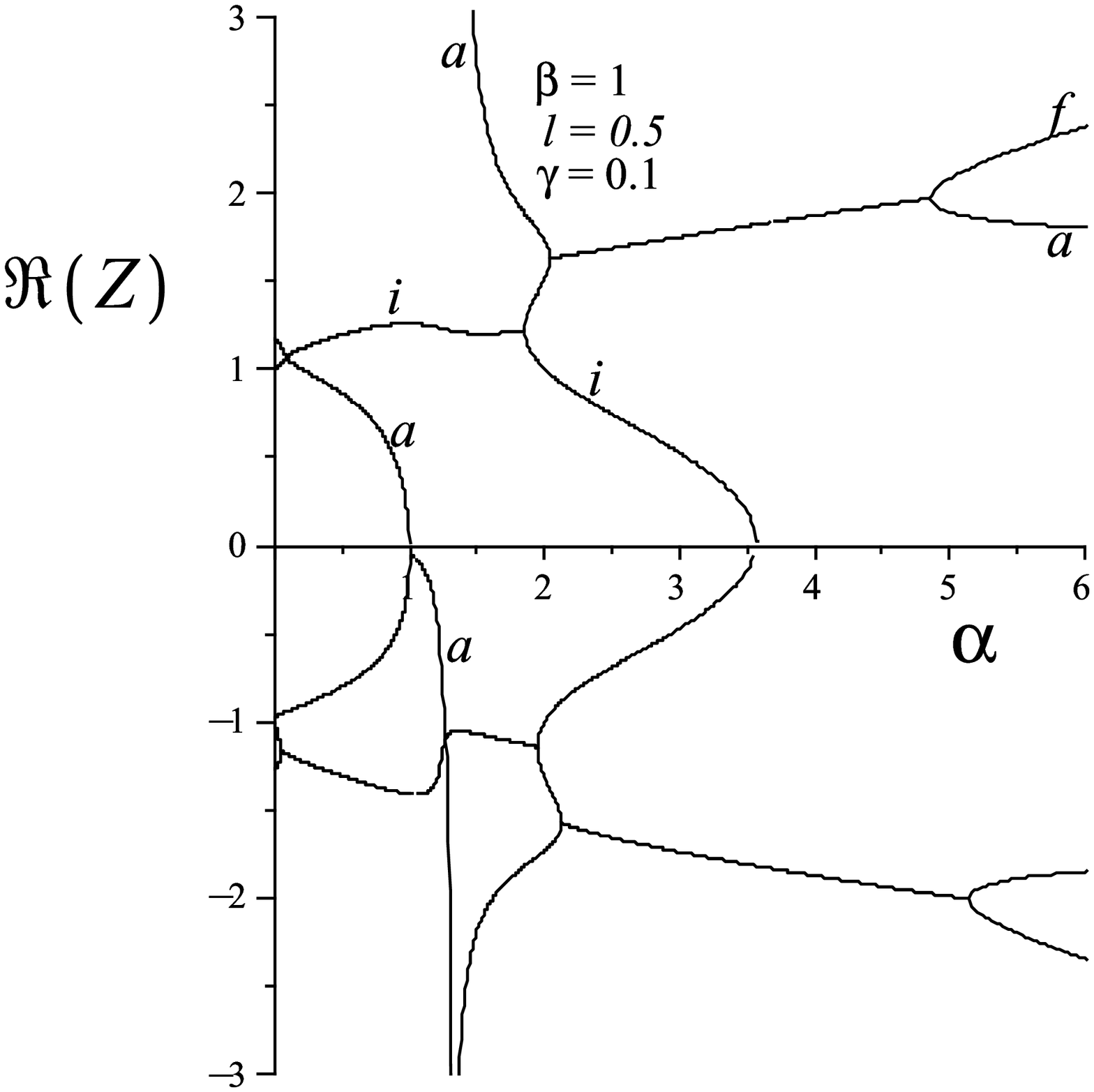}%
\hfill\includegraphics[width=0.43\textwidth]{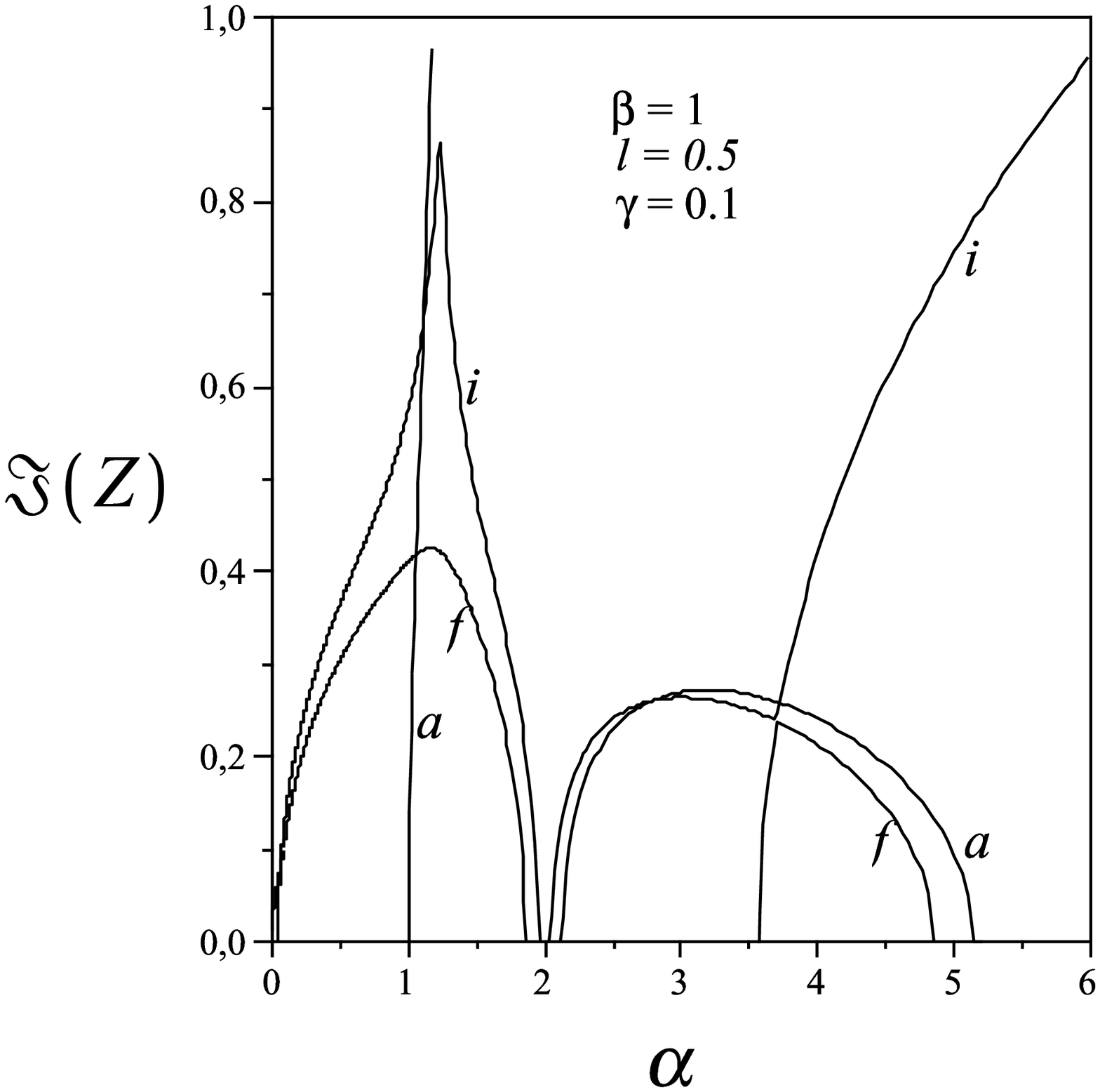}\hspace*{\fill}
\caption{Phase velocities $V_{\mathrm{ph}}=\mathrm{Re}(Z)$ of the
three modes as a function of $\alpha$ for fixed parameters
$\beta=1$, $\phi=\pi/4$, and $\gamma=0.1$ (left picture);
instability rates $\mathrm{Im}(Z)$ of the three modes (right
picture). The labels at the curves correspond to modified fire-hose
waves $(f)$, to isotropic thermal waves $(i)$, and to thermally
anisotropic waves $(a)$.} \label{pol6}
\end{figure*}

To investigate the role of the thermal parameter $\gamma$, we solved
the 6\,th order polynomial equation, choosing more realistic values
of $\gamma<1$. In Fig.(\ref{pol6}), the phase velocities and
instability rates are shown. We observe that\\
a) waves with positive and negative velocities are different and all
 three wave branches become coupled;\\
b) maxima of the instability rate strongly depend on
$\gamma$.\\
For some parameters, all three modes become unstable.
\subsection{Classic fire-hose modes}
To compare our results with the classic fire-hose modes based on the
CGL equations, we use the initial set of equations, omit Eqs.
(\ref{spar}) and (\ref{sper}), and substitute into the others
$S_\|=S_\bot=0$. These CGL equations for incompressible waves
provide two pairs of solutions:
\begin{equation}\label{cglA}
    Z=\pm i\sqrt{A},\ \ A=\frac{(1-\alpha)[\alpha(1+l)-4l]+2\beta(2-\alpha)}
    {\alpha(1+l)-4l} .
\end{equation}
The fire-hose instability condition is that $A>0$. In the parallel
propagation case ($l=1$), this condition is the most familiar case
of $\alpha+\beta<1$. We obtain the same result for $l\ne 1$, if
$\beta=0$. The instability condition is more complete for the
oblique propagation case if $\beta>0$, but this strongly differs
from our results based on the 16--moment equations. For the classic
fire-house instability, we always have that $\mathrm{Re}(\omega)=0$.
In our case, $\mathrm{Re}(\omega)\ne 0$ in most cases due to the
coupling of these modes with other thermal modes. For the classic
modes in the relations, for example, $p_\|'/p_\|=n_1\, B'/B$ and
$p_\bot'/p_\bot=n_2\,B'/B$, the coefficients are $n_1=-2$, $n_2=1$.
In our case, these coefficients are complete functions of all
parameters. The main difference in our case compared to that for the
CGL equations is the appearance of two additional thermal branches,
even if $\gamma=0$.
\section{Conclusion}
To investigate the peculiarities of large-scale wave motions in a
collisionless magnetized plasma, we have applied the 16--moment
transport equations, derived as integrated moments of the kinetic
equations. In earlier similar attempts, the 13--moment equations
were used. However, these equations exclude without any reason the
thermal fluxes and are therefore incomplete.

Anisotropy is the main feature of a collisionless plasma with a
strong magnetic field. In the present study, the pressure anisotropy
was described by the parameter $\alpha$ and the heat fluxes by
$\gamma$. By assuming that $\gamma=0$ we were unable to derive the
13--moment equations, or by assuming that both $\alpha=1$ and
$\gamma=0$ we did not obtain the isotropic MHD case. The 16--moment
equations were in principle different equations. Using these
equations, we illustrated that a wide unstable and stable wave
spectrum in the collisionless anisotropic plasma was possible, even
in the incompressible approximation. If $\gamma\ne0$ (heat fluxes
are present), the waves propagated along and against the magnetic
field at different speeds. This behavior differed from that of the
usual isotropic MHD case. The coupled wave spectrum, including
modified fire-hose modes, strongly depends on the magnetic field
value (parameter $\beta$), pressure anisotropy parameter $\alpha$,
heat flux parameter $\gamma$,  and wave propagation angle $\phi$
with respect to the magnetic field. The deduced instability
increments are rather large. We have derived the general instability
condition for incompressible waves.
\begin{acknowledgements}
We are vary thankful to Bernhard Kliem, Gottfried Mann, and the
referee whose critical comments helped to improve an earlier version
of this paper. The present work has been supported by the German
Science Foundation (DFG) under grant No. 436 RUS 113/931/0-1 (R)
which is gratefully acknowledged.

\end{acknowledgements}

\bibliographystyle{alpha}

\end{document}